\documentclass[%
reprint,
superscriptaddress,
aps,pre,nofootinbib]{revtex4-2}
\usepackage{scrextend}
\usepackage[pdftex]{graphicx}
\usepackage{dcolumn}
\usepackage{bm}
\usepackage{float}
\usepackage{color}
\usepackage{hyperref}
\usepackage{amsmath}
\usepackage{pdfpages}
\usepackage[T1]{fontenc}

\usepackage[utf8]{inputenc}
\usepackage{mathtools}
\usepackage{amssymb} 
\usepackage{fancyhdr}
\usepackage{placeins}
\usepackage{here}

\newcommand{\vect}[1]{\mathbf{#1}}

\newcommand{\rring}{r_0}
\newcommand{\ts}{t_0}
\newcommand{\rme}{\mathrm{e}}
\newcommand{\rmd}{\mathrm{d}}
\newcommand{\lamg}{\frac{\kappa}{\gamma}}
\newcommand{\bs}{\rme^{-\lamg\tau}}

\newcommand{\Deltat}{\tau_\mathrm{s}}
\newcommand{\DeltatE}{\tau_\mathrm{e}}
\newcommand{\tk}{\tilde{k}}
\newcommand{\tl}{\tilde{l}}
\newcommand{\ttil}{\tilde{t}}

\makeatletter
\AtBeginDocument{\let\LS@rot\@undefined}
\makeatother

\begin{document}
	\raggedbottom
\title{Cybloids – Creation and Control of Cybernetic Colloids}

\author{Debasish Saha$^\ddagger$}
\email{debasish.saha@hhu.de}
\affiliation{Condensed Matter Physics Laboratory, Heinrich-Heine-Universit\"at D\"usseldorf, Universit\"atsstra\ss e 1, D-40225 D\"usseldorf, Germany}

\author{Sonja Tarama$^\ddagger$}
\email{sonja.tarama@hhu.de}
\affiliation{Institute for Theoretical Physics II: Soft Matter, Heinrich-Heine-Universit\"at D\"usseldorf, Universit\"atsstra\ss e 1, D-40225 D\"usseldorf, Germany.}
\affiliation{Present address: Laboratory of Biological Computation, College of Life Sciences, Department of Bioinformatics, Ritsumeikan University, 1 Chome-1-1 Nojihigashi, Kusatsu, Shiga 525-0058, Japan}

\author{Hartmut L\"owen}%
\affiliation{Institute for Theoretical Physics II: Soft Matter, Heinrich-Heine-Universit\"at D\"usseldorf, Universit\"atsstra\ss e 1, D-40225 D\"usseldorf, Germany.}

\author{Stefan U. Egelhaaf}
\affiliation{Condensed Matter Physics Laboratory, Heinrich-Heine-Universit\"at D\"usseldorf, Universit\"atsstra\ss e 1, D-40225 D\"usseldorf, Germany}

\date{\today}

\begin{abstract}
Colloids play an important role in fundamental science as well as in nature and technology. They have had a strong impact on the fundamental understanding of statistical physics. For example, colloids have helped to obtain a better understanding of collective phenomena, ranging from phase transitions and glass formation to the swarming of active Brownian particles. 
Yet the success of colloidal systems hinges crucially on the specific physical and chemical properties of the colloidal particles, i.e.~particles with the appropriate characteristics must be available. Here we present an idea to create particles with freely selectable properties. The properties might depend, for example, on the presence of other particles (hence mimicking specific pair or many-body interactions), previous configurations (hence introducing some memory or feedback), or a directional bias (hence changing the dynamics). Without directly interfering with the sample, each particle is fully controlled and can receive external commands through a predefined algorithm that can take into account any input parameters. This is realized with computer-controlled colloids, which we term {\it cybloids} —short for {\it cyb}ernetic col{\it loids}. The potential of cybloids is illustrated by programming a time-delayed external potential acting on a single colloid and interaction potentials for many colloids. Both an attractive harmonic potential and an annular potential are implemented. For a single particle, this programming can cause subdiffusive behavior or lend activity. For many colloids, the programmed interaction potential allows to select a crystal structure at wish.  Beyond these examples, we discuss further opportunities which cybloids offer.
\end{abstract}

\maketitle
\def\thefootnote{$\ddagger$}\footnotetext{These authors contributed equally to this work}

\section{Introduction}

The behaviour of a many-body system is determined by the interactions between its constituents, which can range from atoms, (bio)molecules and colloids to biological cells, animals and humans or even to planets, stars and galaxies. To establish a link between the interactions and the observed behaviour is one of the central tasks of statistical physics\cite{Hafner,Lowen_review}. For a systematic and quantitative experimental test of theoretical predictions, it is crucial that the interactions can be tuned. In atomic systems, the available set of interactions is restricted by the number of atomic species. In contrast, in colloidal systems the interactions can in principle be tuned \cite{Yethiraj_2003,Dzubiella_2009,Schmidt_2022}. To change the colloid-colloid interactions, however, one usually needs to interfere with the sample, e.g., change the chemical composition or structure of the particles or add ingredients such as salts or polymers \cite{Oh_2005}. There are only few systems that allow for external control of the interactions. For example, the interactions of microgel particles and paramagnetic particles are susceptible to changes in temperature \cite{Vogel_review,Bergman_2019,Harrer} and an external magnetic field \cite{Zahn,Keim1,Keim2,Assoud,Horn,Huang_2016}, respectively. However, the range of possible modifications remains quite limited. Irrespective of these limitations, colloids continue to play an important role in fundamental science and have had a particularly large impact on statistical physics. They have helped us to understand collective phenomena, ranging from glass formation
\cite{Pusey_2009,ourbook,Weeks} to phase transitions \cite{Lekkerkerker,Dijkstra_2006,Allahyarov,Li,Weitz_2017} as well as the clustering of active Brownian particles \cite{Buttinoni,Volpe_review}. Furthermore, this understanding has a large impact also on other areas of sciences, ranging from life science, e.g.~the crowding in biological cells \cite{Hoefling_2013} and the swarming of bacteria \cite{Lobaskin_2013,swarming,Ariel_2015,Heffern_2021}, to medicine, e.g. the mechanism leading to cataract \cite{Dorsaz_2011}, and to material sciences, e.g.~the rational design of materials such as the spontaneous assembly of hierarchical structures \cite{Solomon,Manoharan_2015}. Since the success of colloidal systems originates from the specific properties (both chemical and physical) of the colloidal particles, the availability of particles with the appropriate characteristics is of prime importance.\\
\begin{figure*}[tbh]
	\begin{center}
		\includegraphics*[width=0.9\textwidth]{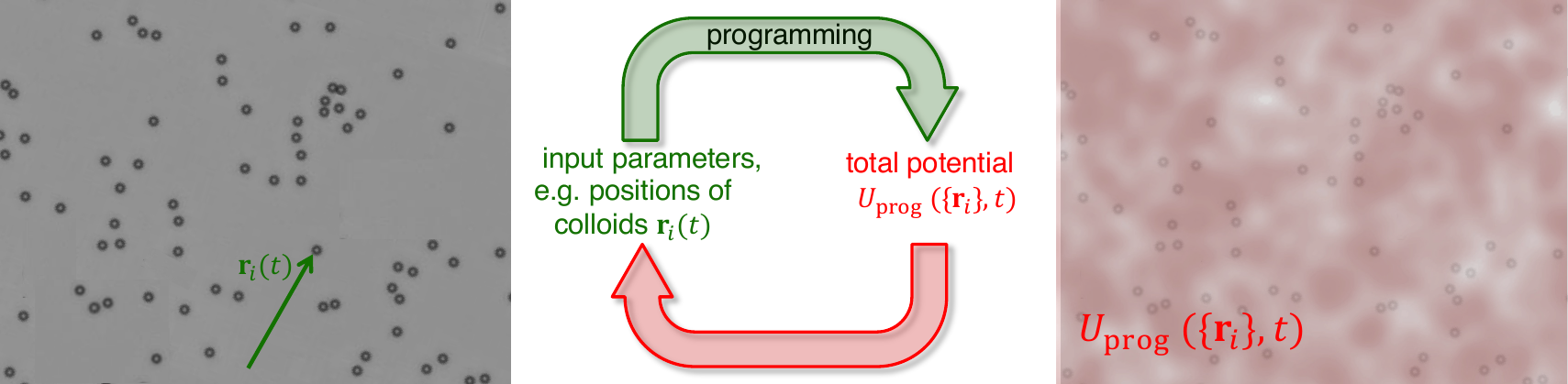}
	\end{center}
	\caption{
		Implementation of cybloids.
		First, particle positions $\vect{r}_i$ are determined from the microscope image. These positions then serve as input parameters for the programmed potential $U_\mathrm{prog}(\{\vect{r}_i\},t)$ which is applied back onto the particles. A (small but finite) time delay $\tau$ naturally arises, due to the time necessary for particle localization and potential calculation. Note that, while here, we constrain ourselves to the simple case of a potential dependent on the particle positions, in principle, any quantity (local density, particle trajectories, etc.) can be used in the programmed potential.
		\label{fig:idea}}
\end{figure*}

\indent Here we present a principle to create colloids with freely selectable properties. The key idea is to expose colloids to an externally created laser-optical potential which is programmed at wish. This allows to program the forces acting on each colloidal particle arbitrarily and thus gives rise to huge flexibility for generating novel colloidal properties. These properties might depend on several parameters, for example, previous configurations (hence introducing some memory or feedback), on the presence of other colloids (hence mimicking specific pair or many-body interactions) or a directional bias (hence changing the dynamics). Each individual particle is fully controlled and can receive commands through a predefined algorithm that can take into account any input parameter without directly interfering with the physical or chemical properties of the particles. We term these computer-controlled colloids
        {\it cybloids} as a portmanteau of  {\it cyb}ernetic col{\it loids}.
The desired properties are implemented by external programming which does neither require to change nor to directly interfere with the sample. This significantly extends the range of properties that can be realised, controlled, investigated and explored to, e.g., design and fabricate previously inaccessible states. Therefore, this method provides new possibilities ranging from novel model systems to new materials.\\

\indent How can colloidal particles be programmed in detail? This is realised using a control circuit (Fig.~\ref{fig:idea}) using time-delayed feedback \cite{Tarama,Fernandez_Rodriguez_2020,Liao_2020,Alvarez_2021,Massana-Cid_2022,Kopp_2023,Kopp_2023_1,Pakpour_2024}. (The details of the experimental setup are given in Section \ref{ch_Materials}.) The particles are imaged using an optical microscope and their positions, $\vect{r}_i(t)$, are followed \cite{crocker1996methods}. The actual and previous particle positions and possibly other parameters are used as input parameters to calculate the total potential, $U_\mathrm{prog}(\vect{r},t)$, that a particle at position $\vect{r}$ should experience at time $t$. The dependence of $U_\mathrm{prog}(\vect{r},t)$ on the input parameters can be freely chosen and determines the characteristics of the cybloids. The potential $U_\mathrm{prog}(\vect{r},t)$ is imposed on the particles by an appropriate light pattern $I(\vect{r},t)$ \cite{Hanes_2009}. This procedure exploits the susceptibility of colloids to external potentials. Here electromagnetic radiation is applied, as in optical tweezers \cite{Ashkin_1986}. 
Whereas tightly focused light beams are used in optical tweezers, extended light patterns are applied to control cybloids \cite{Capellmann_2018}. Modern optical methods allow us to create almost any time-varying light pattern $I(\vect{r},t)$ \cite{Yang_2023} and hence to program a large variety of particle properties, including almost any particle-particle or particle-independent potential, cooperation rule or mode of activity.\\
\indent In Section \ref{ch_feedback_1p}, we illustrate this approach by first implementing a feedback potential for a single colloidal particle. If the cybloid is coupled to a harmonic potential centered around its past position (``attractive'' harmonic potential), this leads to a transient {\it{subdiffusive}} regime as the particle is retracted to its past position. However, if the harmonic potential is inverted, the particle is kicked away from its previous position resulting in an intermediate {\it  ballistic} dynamical regime. In our experiment, we realize such a ``repulsive'' harmonic potential by an annular ring potential \cite{Bell-Davies_2023} centered around the previous particle position. In Section \ref{ch_feedback_multip}, we consider groups of cybloids and program their interaction potential as single or multiple attractive rings around them. By tuning the distance between different rings one can tune the crystalline lattice  and generate exotic crystallites such as ones dominated by a square structure in two dimensions which is relatively uncommon for single-component systems.
We compare our experimental results with theory and Brownian dynamics computer simulations and we find good agreement.

\section{Materials and methods}\label{ch_Materials}
\subsection{Sample preparation and experimental methods}\label{ch_Materials_experiment}
We use colloidal suspensions of sulfate latex beads (IDC) with radius $R = 1.4~\mu$m in heavy water (D$_{2}$O). Due to the larger density of heavy water ($1.11$~ g/ml), the particles ($1.05$~g/ml) are pushed to the top by gravity as well as by radiation pressure. The suspension is stirred together with ion exchange resin to reduce the salt content. Although no further effort is taken to keep the salt concentration low, this is sufficient to avoid particles sticking to the glass surface. Subsequently, the suspension is loaded into a sample cell. The cell is constructed from a microscope slide and a cover glass (number \#$1$) separated by two cover glasses (number \#$0$) used as spacers that are fixed using UV glue \cite{Jenkins_2008}. This results in a narrow capillary which is sealed with glue after filling.
\indent
The colloidal particles are observed using an inverted microscope (Eclipse TE2000-U, Nikon) equipped with a $60\times$ objective. Images are recorded using a CMOS camera with USB interface (MAKO U-130, AVT) and the particle locations $\{\vec{r}_i(t)\}$ are determined in real-time with a LabVIEW subroutine (IMAQ Particle Analysis Report VI) that is based on a standard particle tracking algorithm.
Light patterns $I(\vect{r},t)$ are created with a holographic optical tweezers setup \cite{Hanes_2009} that is combined with the inverted microscope. It is based on a SLM i.e.~spatial light modulator (LCR-2500, Holoeye), 2-D galvanometer-mounted mirrors (Quantum Scan 30, Nutfield Technology Inc.) and a diode-pumped solid-state laser (wavelength $\lambda = 532$~nm, Ventus 532-1500, Laser Quantum). A control software based on LabVIEW and MATLAB handles the image acquisition, particle tracking and hologram generation. The holograms are calculated using an iterative algorithm, the Gerchberg-Saxton algorithm \cite{Gerchberg_1972} and uploaded to the SLM. Finally, the light field is obtained by the speckle pattern created by the SLM. The particle tracking and light field generation naturally leads to a time delay $\tau$ between the measured particle positions $\{\vec{r}_i(t)\}$ and the introduction of the corresponding light field $I(\vect{r},t)$. The light field is kept the same while the next one is calculated, leading to discrete potential update times $n\tau$ (where \mbox{$n=0,1,2,\ldots$}). Here, the delay time $\tau= 0.2\mathrm{s}$ is much smaller than the Brownian time 
$\tau_\mathrm{B} = R^2/D_0 \approx 40\mathrm{s}$ 
where $D_0 \approx 0.05~\mu\mathrm{m}^2/\mathrm{s}=$ is the diffusion coefficient without an external potential but in the presence of the confining glass plates. This implies that the feedback can be applied very quickly in an almost instantaneous manner.
\subsection{Langevin equation of motion and computer simulations}\label{ch_Materials_ana_simu}
We model the time-dependent trajectory $\vect{r}(t)=(x(t),y(t))$ of a single colloid in two spatial dimensions using Brownian dynamics. We describe the particle motion by the time-delayed over-damped Langevin equation\cite{Kuechler1992,Giuggioli2016}
\begin{equation}
\frac{\rmd\vect{r}}{\rmd t}=\frac{1}{\gamma}\vect{F}_\mathrm{prog}(\vect{r}-\vect{r}_\mathrm{p})+\frac{1}{\gamma}\vect{f}(t)\,,\label{eq_eom}
\end{equation}
where $\gamma$ denotes the friction coefficient and $\vect{F}_\mathrm{prog}(\vect{r}-\vect{r}_\mathrm{p})$ is the systematic force the particle experiences at its position $\vect{r}(t)$ due to the feedback potential centered around its past position $\vect{r}_\mathrm{p}$.
$\vect{f}(t)$ is a Gaussian random force that describes the thermal motion of the particle and is characterized by its first two moments $\langle \vect{f}(t)\rangle=0$ and \mbox{$\langle \vect{f}(t)\vect{f}(t')\rangle=2dD_0\gamma^2\delta\left(t-t'\right)$}, where $\langle\ldots\rangle$ indicates an average over different realizations of the noise, $D_0$ is the diffusion coefficient of the particles and $d=2$ gives the spatial dimension.
The feedback force
\begin{equation}
\vect{F}_\mathrm{prog}(\vect{r})=-\nabla U_\mathrm{prog}(\vect{r})\,
\end{equation}
is derived from the programmed potential $U_\mathrm{prog}(\vect{r})$. We use two types of potentials: a harmonic attractive potential  
\begin{equation}
U(\vect{r})=\frac{1}{2}\kappa \vect{r}^2\label{eq_Vext_harm}
\end{equation}
and a Gaussian ring potential
\begin{align}
V(\vect{r})&=V_0\ \rme^{-\frac{\left(|\vect{r}|-\rring\right)^2}{2b^2}}\,.\label{eq_Vext_Gauss}
\end{align}
As in the experiment, we define times $n\tau$ with \mbox{$n=0,1,2,\ldots$} at which the position of the particle is evaluated to update the past position $\vect{r}_\mathrm{p}$ entering into the potential. 
The programmed potential $U_\mathrm{prog}(\vect{r})$ is then given by eq.~(\ref{eq_Vext_harm}) and eq.~(\ref{eq_Vext_Gauss}) respectively and centered around the past particle position
\begin{equation}
\vect{r}_\mathrm{p}=\vect{r}((k-1)\tau)
\end{equation}
for the time interval $k\tau\le t < (k+1)\tau$, such that the last multiple of $\tau$ smaller than time $t$ is $k\tau$ and the feedback potential introduced at this point corresponds to the particle positions a delay time $\tau$ prior to this which is $(k-1)\tau$. The feedback potential is kept the same until the next potential update at time $(k+1)\tau$. Consequently, the past position used in the feedback term has a time shift of between $\tau$ and $2\tau$ to the present time $t$.

For the case of multiple particles, the programmed potential $U_\mathrm{prog}$ is composed of contributions of the form of eq.~(\ref{eq_Vext_Gauss}) centered around all past particle positions $r_{\mathrm{p},j}$. Here, $r_{\mathrm{p},j}=\vect{r}_j((k-1)\tau)$ indicates the past position of the $j$th particle. Further, we introduce an additional force to describe the direct repulsive steric interactions between the particles. This part is modelled via a smooth Weeks-Chandler-Anderson (WCA) pair-potential which takes the form
\begin{align}
V_\mathrm{WCA}(r)=\begin{cases}
4\epsilon\left[\left(\frac{\sigma}{r}\right)^{12}-\left(\frac{\sigma}{r}\right)^6\right]+\epsilon\,, &
r\le 2^{\frac{1}{6}}\sigma\,,\\
0 &\text{else}\,,
\end{cases}
\end{align}
where $\sigma$ is the particle diameter, $r$ is the distance between the two particles and $\epsilon$ sets the interaction strength. Thus, the equation of motion in the multiple-particle case reads
\begin{align}
\frac{\rmd \vect{r}_i}{\rmd t}=\frac{1}{\gamma}\vect{f}_i(t)
&-\frac{1}{\gamma}\nabla_i\sum_{\substack{j=1\\ j\ne i}}^{N}\vect{V}_\mathrm{WCA}(|\vect{r}_i-\vect{r}_j|)\nonumber\\
&+\frac{1}{\gamma}\sum_{j=1}^{N}\vect{F}_\mathrm{prog}(\vect{r}_i-\vect{r}_{p,j})\label{eq_eom_multi_p}
\end{align}
such that every particle $i$ is experiencing a fluctuating Brownian force $\vect{f}_i$, direct interaction via the WCA potential with all particles $j$ but itself ($j\ne i $) and feedback forces including the self-term $j=i$.\\
\indent
Brownian dynamics simulations are performed based on eqs.~(\ref{eq_eom}) and (\ref{eq_eom_multi_p}) with an explicit Euler forward integration scheme with a finite time step $\Delta t=10^{-4}\mathrm{s}$ (all other parameters are given in the subsequent figure captions explicitly). For the simulations, the system is first equilibrated without the programmed potential (pure diffusive motion). Then, the feedback potential is introduced and the dynamics again equilibrated for $\sim 100\mathrm{s}$.

\section{Feedback-driven dynamics of a single colloidal particle}\label{ch_feedback_1p}
First we investigate the effect of a feedback potential on the dynamics of a single colloidal particle. Here, we study simple forms of the programmed potential, given by a harmonic trap (eq.~(\ref{eq_Vext_harm})) and an annular ring potential (eq.~(\ref{eq_Vext_Gauss})). In general, any interaction potential can be implemented. We shall first present an analytic solution for a harmonic trap and then turn to experiment and simulation.

\subsection{Analytic solution for a harmonic feedback trap} \label{ch_ana_harmonic_trap}
\begin{figure}[tb]
	\centering
	\includegraphics[width=0.99\columnwidth]{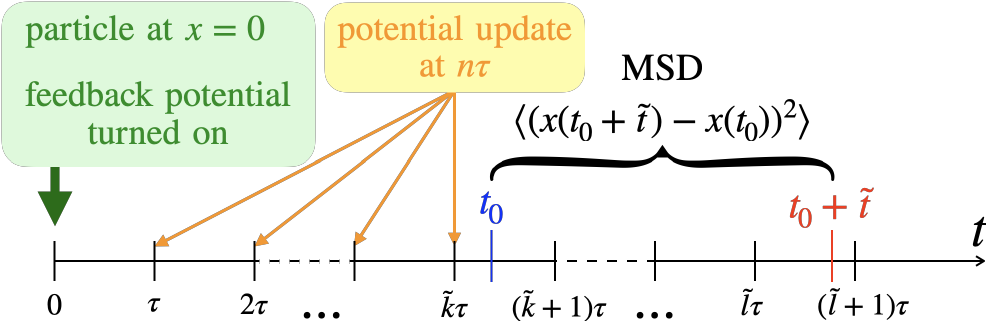}
	\caption{Timeline of the system. The feedback is turned on at $t=0$ with the particle at $x=0$ and the feedback potential centered at this position. The MSD is calculated between two later times $t_0$ and $t_0+\ttil$ that lie in the intervals $\tk\tau \le\ts <(\tk+1)\tau$ and $\tl\tau\le t_0+\ttil< (\tl+1)\tau$ with $\tk,\tl\in 0,1,2,\ldots$ 
		determined by these relations (where $\tk\leq\tl$). The feedback potential is updated at multiples of the delay time $\tau$. The equation of motion is solved piece-wise for the intervals $k\tau\le t< (k+1)\tau$, i.e.~between two feedback potential update points.}\label{fig_sketch_ana}
\end{figure}
\FloatBarrier
In the following we derive an analytic expression for the mean squared displacement (MSD) of a single particle in a two-dimensional harmonic feedback potential. For the harmonic potential, the programming force is given by
\begin{align}
\vect{F}(\vect{r}-\vect{r}_\mathrm{p})= -\kappa\left(\vect{r}-\vect{r}_\mathrm{p}\right)
\end{align}
with the actual and past particle positions $\vect{r}=\vect{r}(t)$ and \mbox{$\vect{r}_\mathrm{p}=\vect{r}((k-1)\tau)$} respectively. Here, \mbox{$k\in 0,1,2,\ldots$} is determined by the relation $k\tau\le t < (k+1)\tau$, i.e.~$k\tau$ is the last update time of the feedback potential before time $t$.
For the harmonic potential, the $x$ and $y$ directions of motion decouple. In the following, we thus continue on to solve the one-dimensional equation of motion
\begin{equation}
\frac{\rmd x}{\rmd t}=-\lamg\left(x(t)-x_\mathrm{p}\right)+\frac{1}{\gamma}\,\mathrm{f}(t)\,,\label{eq_eom_harmonic}
\end{equation}
with $x_\mathrm{p}=x((k-1)\tau)$. From this the two-dimensional MSD, follows as
\begin{align}
\langle &\left( \vect{r}(\ts+\ttil)-\vect{r}(\ts)\right)^2\rangle\nonumber\\&=\langle \left( x(\ts+\ttil)-x(\ts)\right)^2\rangle+\langle \left( y(\ts+\ttil)-y(\ts)\right)^2\rangle\nonumber\\&=2\langle \left( x(\ts+\ttil)-x(\ts)\right)^2\rangle\,.\label{eq_MSD}
\end{align}
Here, we define the MSD as the mean squared distance the particle travels within a time $\ttil$ starting from a fixed reference time $\ts$. It is thus given as the mean squared distance traveled between the fixed times $\ts$ and $\ts+\ttil$, while averaging over different realizations of the thermal noise with $\langle\ldots\rangle$ denoting the thermal noise average.

We solve eq.~(\ref{eq_eom_harmonic}) on the intervals between feedback potential updates $k\tau\le t < (k+1)\tau$. Within these time intervals $x_\mathrm{p}$ does not change, so we use the ansatz $x(t)=x_0(t)\,\rme^{-\lamg t}$ to obtain the solution for the particle position
\begin{align}
x(t)=&\rme^{-\lamg t}\int_{k\tau}^{t}\rmd t' \frac{\mathrm{f}(t')}{\gamma}\rme^{\lamg t'} +x\left(k\tau\right)\rme^{-\lamg\left(t-k\tau\right)}\nonumber\\
&+x\left(\left(k-1\right)\tau\right)\left(1-\rme^{-\lamg\left(t-k\tau\right)}\right)\label{eq_int}
\end{align}
where the integration constant is set by $x(t=k\tau)=x(k\tau)$.
This solution still depends on the previous positions of the particle. For $t\to (k+1)\tau$, we obtain the recursive formula for the particle position at the feedback potential update times
\begin{align}
x((k+1)\tau)=&\rme^{-\lamg (k+1)\tau}\int_{k\tau}^{(k+1)\tau}\rmd t' \frac{\mathrm{f}(t')}{\gamma}\rme^{\lamg t'}+
x\left(k\tau\right)\rme^{-\lamg\tau}\nonumber\\
&+x\left(\left(k-1\right)\tau\right)\left(1-\rme^{-\lamg\tau}\right)\,.
\label{eq_xkp1_rec}
\end{align}

Now, from eq.~(\ref{eq_xkp1_rec}), we can see that the thermal noise within the last delay time (from $k\tau$ to $(k+1)\tau$) enters with the term
\begin{align}
    \rme^{-\lamg (k+1)\tau}\int_{k\tau}^{(k+1)\tau}\rmd t' \frac{\mathrm{f}(t')}{\gamma}\rme^{\lamg t'}\nonumber
\end{align}
while previous positions (that depend on the noise integrals at earlier times) enter with additional constant prefactors. We thus use the ansatz 
\begin{equation}
x(k\tau)=\rme^{-\lamg k \tau}\sum_{n=0}^{k-1}c_n\int_{(k-n-1)\tau}^{(k-n)\tau}\rmd t' \frac{\mathrm{f}(t')}{\gamma}\rme^{\lamg t'}\label{eq_ansatz_xktau}
\end{equation}
for the particle position at the update times $k\tau$, where $c_n$ are (parameter-dependent) constants. Here, we cut the sum at $k-1$, i.e.~the last interval entering into the sum is for $t=0$ to $t=\tau$. This demands that the recursion according to eq.~(\ref{eq_xkp1_rec}) ends at
\begin{align}
x(\tau)=&\rme^{-\lamg \tau}\int_{0}^{\tau}\rmd t' \frac{\mathrm{f}(t')}{\gamma}\rme^{\lamg t'}+
x\left(0\right)\rme^{-\lamg\tau}\nonumber\\&+x\left(-\tau\right)\left(1-\rme^{-\lamg\tau}\right)\,.
\label{eq_xkp1_rec_k0}
\end{align} 
We thus prescribe $x(0)=0$ and $x(-\tau)=0$ at these two times, i.e., the particle starts from the position $x(0)=0$ and the first feedback potential is centered at this position.

Next, we plug eq.~(\ref{eq_ansatz_xktau}) into both sides of eq.~(\ref{eq_xkp1_rec}). As the noise is $\delta$-correlated, the integrals for different time intervals are completely independent of each other and should as such have matching prefactors on both sides of the equation. We thus compare the respective prefactors of the integrals to find
\begin{align}
\begin{cases}
    c_0&=1\\
    c_1&=c_0\\
    c_n&=c_{n-1}+\rme^{2\lamg \tau}\left(1-\rme^{-\lamg\tau}\right)c_{n-2}
\end{cases}
\,.
\end{align}
Finally, we solve this recurrence equation for $c_n$ via the characteristic root technique to obtain
\begin{equation}
c_n=\rme^{n\lamg\tau}\frac{1}{2-\rme^{-\lamg\tau}}\left(1-\left(\rme^{-\lamg\tau}-1\right)^{n+1}\right)
\end{equation}
and arrive at the full solution for the particle position
\begin{align}
x(t)=&\rme^{-\lamg t}\int_{k\tau}^{t}\rmd t'  \frac{\mathrm{f}(t')}{\gamma}\rme^{\lamg t'}+\sum_{n=0}^{k-1}c_n\int_{\left(k-n-1\right)\tau}^{\left(k-n\right)\tau}\rmd t' \frac{\mathrm{f}(t')}{\gamma}\rme^{\lamg t'}\rme^{-\lamg t}\nonumber\\
&+\sum_{n=0}^{k-2}c_n\int_{\left(k-n-2\right)\tau}^{\left(k-n-1\right)\tau}\rmd t' \frac{\mathrm{f}(t')}{\gamma}\rme^{\lamg t'}\rme^{\lamg \tau}\left(\rme^{-\lamg k\tau}-\rme^{-\lamg t}\right)\,.
\end{align}

The mean squared displacement (MSD) for $\tk\tau \le\ts <(\tk+1)\tau$ and $\tl\tau\le t_0+\ttil< (\tl+1)\tau$ (and $\tk\leq\tl$), cf.~Fig.~\ref{fig_sketch_ana}, i.e.~the mean squared distance the particle travels between two fixed times $\ts$ and $\ts+\ttil$ while averaging over different realizations of the thermal noise, is then given by

\begin{widetext}
\begin{align}
	\langle& \left( x(\ts+\ttil)-x(\ts)\right)^2\rangle\frac{\kappa/\gamma}{D_0}
	\nonumber\\ &=2-\rme^{-2\lamg\DeltatE}-\rme^{-2\lamg\Deltat}+\left(1-\rme^{-2\lamg\tau}\right)
	\Bigg[
	A(\tl-1,\tl-1)\rme^{-2\lamg\DeltatE}+A(\tl-2,\tl-1)\rme^{-\lamg \tau}\rme^{-\lamg\DeltatE}\left(1-\rme^{-\lamg\DeltatE}\right)\nonumber\\ &
	+A(\tl-1,\tl-2)\rme^{\lamg\tau}\rme^{-\lamg\DeltatE}
	\left(1-\rme^{-\lamg\DeltatE}\right)+A(\tl-2,\tl-2)\left(1-\rme^{-\lamg\DeltatE}\right)^2
	+A(\tk-1,\tk-1)\rme^{-2\lamg\Deltat}\nonumber\\ &+A(\tk-2,\tk-1)\rme^{-\lamg \tau}\rme^{-\lamg\Deltat}\left(1-\rme^{-\lamg\Deltat}\right)+A(\tk-1,\tk-2)\rme^{\lamg\tau}\rme^{-\lamg\Deltat}
	\left(1-\rme^{-\lamg\Deltat}\right)\nonumber\\ &\qquad\qquad+A(\tk-2,\tk-2)\left(1-\rme^{-\lamg\Deltat}\right)^2\Bigg]\nonumber\\
&+\rme^{-\lamg \ttil}\Bigg[ \left(1-\rme^{-2\lamg\Deltat}\right)\delta_{\tk\tl}+A(\tk-1,\tl-1)\left(1-\rme^{-2\lamg\tau}\right)\rme^{-2\lamg\Deltat}+A(\tk-2,\tl-1)\left(1-\rme^{-2\lamg \tau}\right)\rme^{-\lamg\Deltat}\rme^{-\lamg \tau}\left(1-\rme^{-\lamg\Deltat}\right)\nonumber\\
&\qquad\qquad +c_{\tl-\tk-1}\left(1-\rme^{-2\lamg\Deltat}\right) \Bigg]\nonumber\\
&+A(\tk-1,\tl-2)\left(1-\rme^{-2\lamg\tau}\right)\rme^{\lamg\tau}\rme^{-\lamg\Deltat}
\left(\rme^{-\lamg\left(\tl-\tk\right)\tau}-\rme^{-\lamg\left(\Deltat +\ttil\right)}\right)\nonumber\\
&+A(\tk-2,\tl-2)\left(1-\rme^{-2\lamg\tau}\right)\left(1-\rme^{-\lamg\Deltat}\right) \left(\rme^{-\lamg\left(\tl-\tk\right)\tau}-\rme^{-\lamg\left(\Deltat+\ttil\right)}\right)+c_{\tl-\tk-2}\ \rme^{\lamg\tau}\left(1-\rme^{-2\lamg\Deltat}\right)\left(\rme^{-\lamg\left(\tl-\tk\right)\tau}\rme^{\lamg\Deltat}-\rme^{-\lamg \ttil}\right)\,.\label{eq_MSD}
\end{align}
\end{widetext}

\begin{figure}[tbh]
	\centering
\includegraphics[width=0.99\columnwidth]{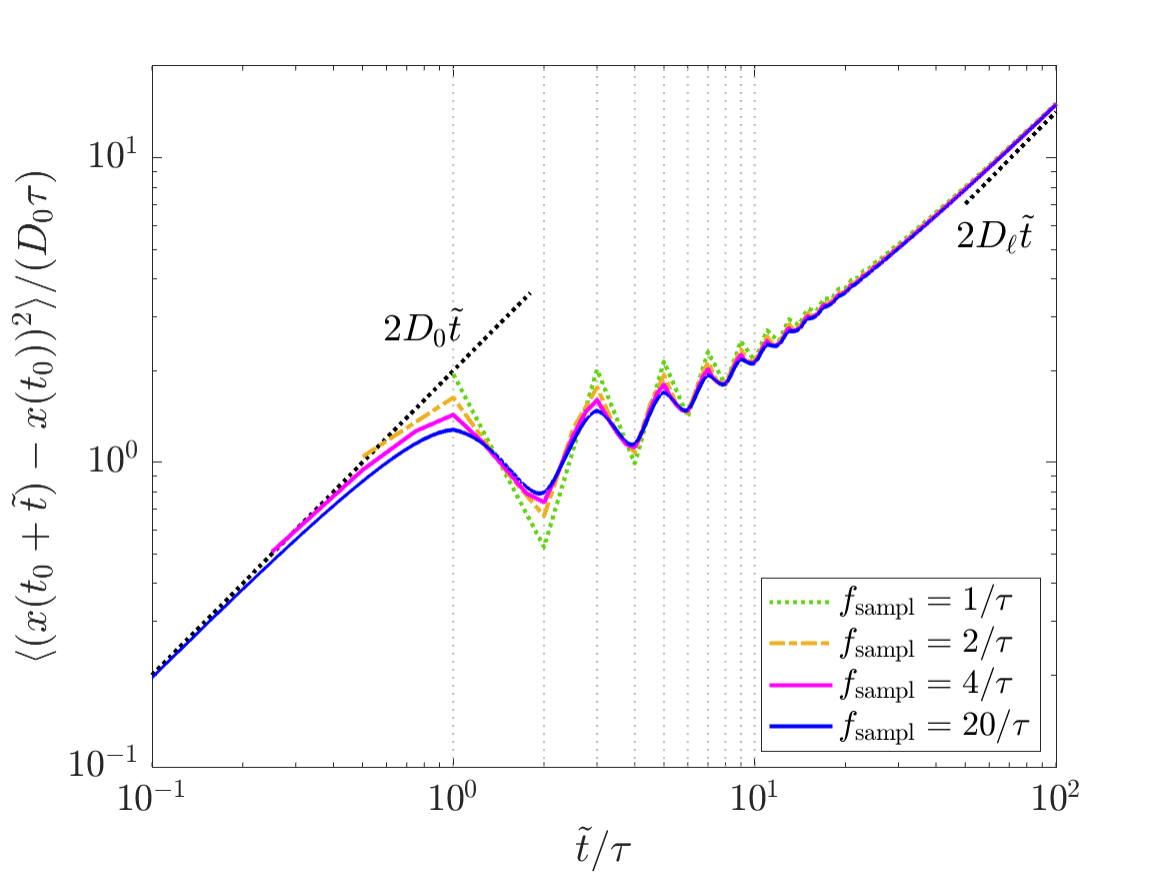} 
\caption{Analytic result for the MSD of a single particle in an attractive harmonic feedback potential (\mbox{$\lamg \tau=2.0$}) and different sampling rates $f_\mathrm{sampl}$ (= number of measurements of the particle position per delay time $\tau$) of the particle position. We find a decrease in the MSD compared to free diffusion (as indicated by the black line $2D_0\ttil$ at short times) and oscillations whose period is set by $2\tau$ which arise as the particle is being pulled back to its past position. At long times the particle shows diffusion with a reduced diffusion constant $D_\ell<D_0$. Discrepancies appear in the MSD for small sampling rates as these do not cover the full average over $t_0$. A high sampling rate is thus necessary to obtain the ``correct'' MSD.}\label{fig_AnaMSD}
\end{figure}

\noindent
Here, $c_n=0$ for $n<0$ and $\Deltat=\ts-\tk\tau$, $\DeltatE=\ts+\ttil-\tl\tau$ are the time offsets to the respective last potential update  ($\Deltat$, $\DeltatE \in [0,\tau)$). Finally, the abbreviation $A(i,j)$ is defined as
\begin{align}
&A(i,j)=\sum_{n=0}^{i}c_nc_{n-i+j}\,\rme^{-2\lamg n\tau}\label{eq_Akm0}
\end{align}
which can be solved as a geometric series as
\begin{align}
&A(i,j)=\frac{\rme^{\lamg\left(j-i\right)\tau}}{\left(\bs-2\right)^2}\Bigg\{\frac{\left(\bs-1\right)^{2i+2}-1}{\left(\bs-2\right)\bs}\left(\bs-1\right)^{j-i+2}\nonumber\\
&-\frac{\left(\bs-1\right)^{i+1}-1}{\bs-2}\left[1+\left(\bs-1\right)^{j-i}\right]\left(\bs-1\right)+i+1\Bigg\}\,.\label{eq_Akm}
\end{align}
Eq.~(\ref{eq_MSD}) gives the mean squared distance that a particle travels between the times $\ts$ and $\ts+\ttil$.
For sufficiently large $\ts$, i.e., large $\tk$, the effect of the initial condition disappears. In this limit, the two terms $(\bs-1)^{2i+2}$ and $(\bs-1)^{i+1}$ in eq.~(\ref{eq_Akm}) vanish (as long as $\bs<2$).
Note, however, that eq.~(\ref{eq_MSD}) not only depends on the time difference $\ttil$ as well as the number of potential renewals $\tl-\tk$ during that time but also on the offsets $\Deltat=\ts-\tk\tau$ and $\DeltatE=\ts+\ttil-\tl\tau$ that the times $\ts$ and $\ts+\ttil$ have to the respective last renewal point of the potential. This is due to the discrete potential updates which cause the system to lose its invariance under translation in time (it is invariant only for multiples of the delay time $\tau$). For measurements of the MSD, the resulting curve may thus vary depending on the protocol used. In particular, to obtain a full time average over $t_0$, the sampling of the particle position must be more frequent than the feedback update. Otherwise, i.e.~for low sampling rates (= number of measurements of the particle position per delay time $\tau$), the resulting MSD curve will not only be ``rough'' due to the small number of data points but actually differ from the full average over $t_0$, see Fig.~\ref{fig_AnaMSD}

Further, we have not made any assumptions with respect to the sign of $\kappa$. The above expressions hold for both attractive ($\kappa>0$) and repulsive potentials ($\kappa<0$). However, for large negative values of $\kappa$, the particle continues to be accelerated away from its past position with increasing forces and the MSD diverges exponentially. An easy way to see this is from eq.~(\ref{eq_xkp1_rec}). Subtracting $x(k\tau)$ from this equation, we find that the distance the particle travels between two potential updates
\begin{align}
x((k+1)\tau)-x(k\tau)=&
\left[ x\left(k\tau\right)-x\left(\left(k-1\right)\tau\right)\right]\left(\rme^{-\lamg\tau}-1\right)\nonumber\\
&+\rme^{-\lamg (k+1)\tau}\int_{k\tau}^{(k+1)\tau}dt' \frac{\mathrm{f}(t')}{\gamma}\rme^{\lamg t'} \label{eq_xkp1}
\end{align}
grows with a scaling factor of $\rme^{-\lamg\tau}-1$ (on average, the additional noise term averages to zero). If this scaling factor is larger than $1$ the distance increases exponentially for repeated iterations. This implies that $\rme^{-\lamg\tau}$ should be smaller than $2$ to ensure convergence. 
We therefore constrain ourselves to potential strengths that fulfill \mbox{$\rme^{-\lamg\tau}<2$}. In this case, at long times the particle dynamics becomes diffusive with the long-time diffusion coefficient
\begin{align}
\frac{D_\ell}{D_0}&=\frac{1}{D_0} \lim_{\ttil\to \infty}\frac{\langle \left( x(\ts+\ttil)-x(\ts)\right)^2\rangle}{2\ttil}
=\frac{1-\rme^{-2\lamg\tau}}{2\lamg\tau\left(2-\rme^{-\lamg\tau}\right)^2}
\end{align}
which only depends on the strength of the external potential $\kappa$ and the delay time $\tau$.

\begin{figure*}[tbh]
	\centering
	\includegraphics[width=6in]{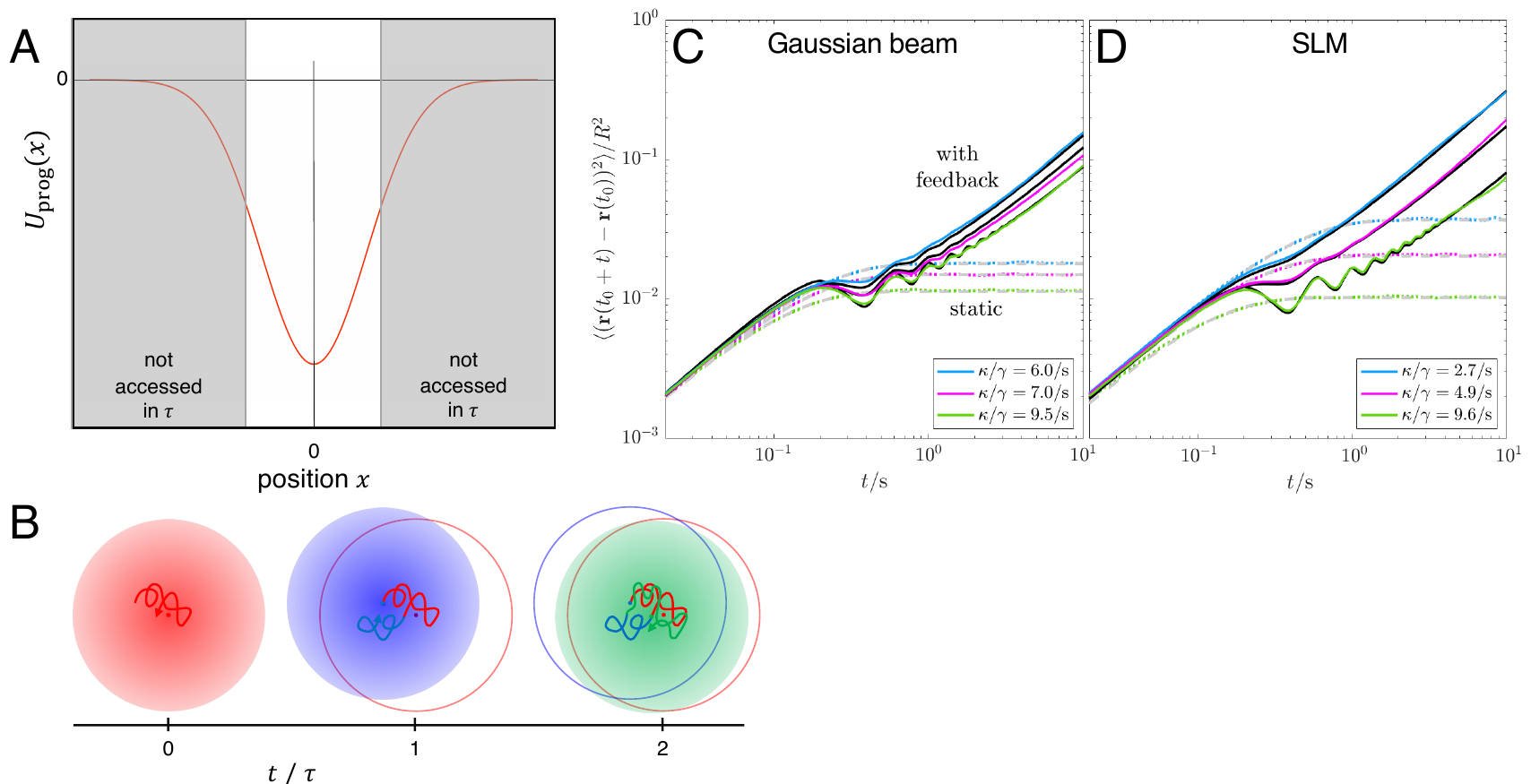}
	\caption{(A) One dimensional cut through the applied attractive potential. (B) The harmonic potential is centered around the position that the particle occupied a time $\tau$ earlier. Trajectories within the respective potential are indicated by the same color.
		The potential is kept the same for a duration $\tau$ while the next potential is calculated. As a result, the starting point of the coloured trajectory becomes the centre of the next programmed potential and the particle is pulled back to this previous position. (C,D) MSD as a function of time $t$ obtained while applying a static harmonic potential (colored dotted lines (experimental data), dashed gray (analytic results)) and a harmonic feedback potential (coloured solid lines (experimental data), black lines (analytic results)) which was created by (C) a Gaussian beam and (D) the SLM. [Parameters of the analytic curves: $D_0=0.026R^2/\mathrm{s}$, $\tau=0.2\mathrm{s}$, values of $\kappa/\gamma$ given in legend]}
	\label{fig:attractive}
\end{figure*}

For an attractive potential ($\kappa>0$), the feedback potential leads to a decrease in the MSD compared to free diffusion as the particle is pulled back towards its past position, see Figs.~\ref{fig_AnaMSD},\ref{fig:attractive}. Additionally, oscillations of period $2\tau$ appear that are linked to the potential update: After a potential update at time $t=0$, the particle diffuses in a non-moving harmonic potential until the next potential update a delay time $\tau$ later. The new potential (at $t=\tau$) is then centered around the position the particle had at the previous potential update (at $t=0$), i.e.~the particle experiences a force drawing it back towards its past position, leading to a reduction in the MSD. At the next potential update (at $t=2\tau$), the particle is then drawn towards its position at time $t=\tau$, driving the particle away from the initial starting position and thus leading to an increase in the MSD (see also Fig.\ref{fig:attractive}B). The repetition of this process leads to the observed oscillations while the additional spatial diffusion leads to overall diffusive behaviour at long times with diffusion coefficient $D_\ell<D_0$. This long time diffusive motion can be understood as the diffusive motion of the feedback potential center.

\begin{figure}[tb]
\centering
\includegraphics[width=0.99\columnwidth]{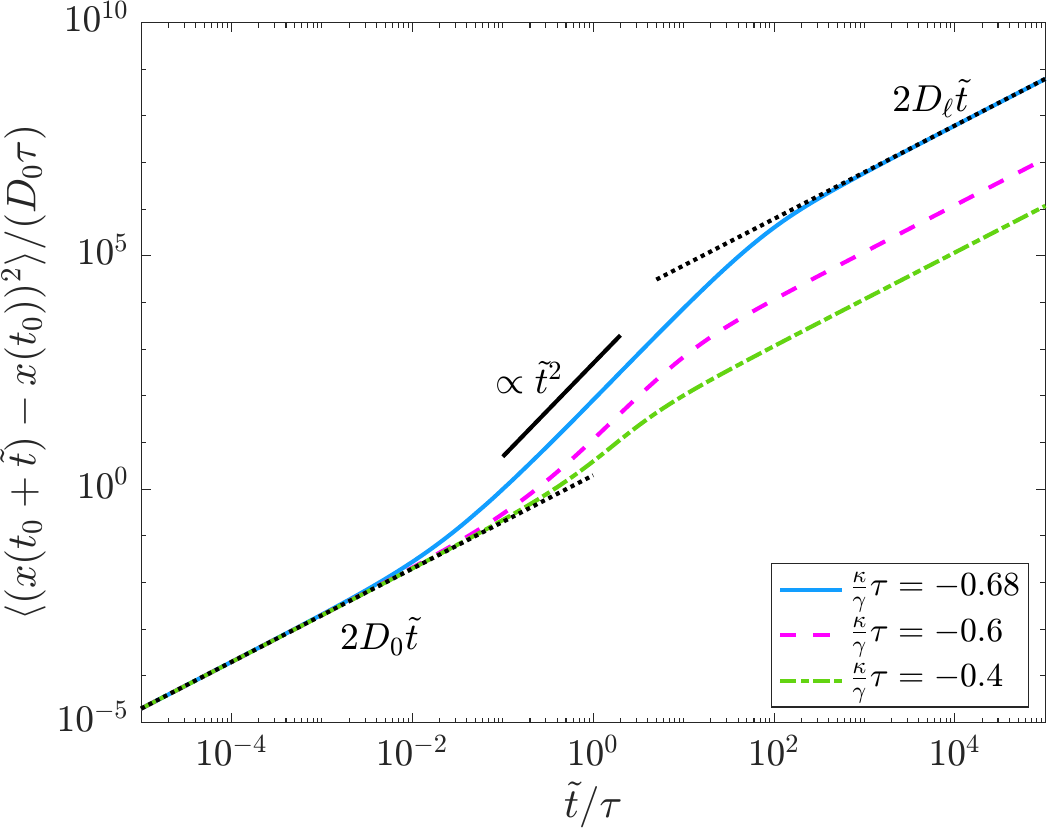}
\caption{Analytic result for the MSD in repulsive harmonic feedback potentials of varying strength (average taken over $t_0$). The repulsive feedback potential, pushing the particle away from its past position, leads to propulsion of the particle as indicated by the ballistic $\propto \tilde{t}^2$ regime of the MSD at intermediate times. Spatial diffusion additionally leads to changes in the angle between current and past position, causing the motion to become diffusive again on long time scales.}\label{fig_AnaMSD_rep}
\end{figure}

For a weakly repulsive potential ($\kappa<0$, $\rme^{-\lamg\tau}<2$), we find a ballistic ($\propto \ttil^2$) regime at medium times between two diffusive regimes with diffusion constants $D_0$ and $D_\ell>D_0$, see Fig.~\ref{fig_AnaMSD_rep}. At short times, the particle dynamics is dominated by diffusion. On intermediate time scales, the introduced feedback potential leads to an effective particle propulsion, as the particle is pushed away from its past position. Due to spatial diffusion, the angle between current and past position changes, causing the motion to become diffusive again on long time scales but now with a larger diffusion coefficient $D_\ell>D_0$. 
Getting closer to the divergence point, the transition from propulsion to long-time diffusion shifts to longer times. This is accompanied by the divergence of the long-time diffusion coefficient as
\begin{align}
\frac{D_\ell}{D_0}=& \frac{3}{2\ln(2)}\frac{1}{\left(p-2\right)^2}+\frac{8\ln(2)-3}{4\left(\ln 2\right)^2}\frac{1}{\left(p-2\right)}\nonumber\\&+\frac{6+8\left(\ln 2\right)^2-13\ln(2)}{16\left(\ln(2)\right)^3}+\mathcal{O}\left(p-2\right)\,,
\end{align}
where $p=\rme^{-\lamg\tau}$.

It should also be noted that, as the memory of previous displacements decays with a factor $\rme^{-\lamg\tau}-1$ over one decay time (see eq.~(\ref{eq_xkp1})), this decay becomes increasingly slow towards the divergence point and the dynamics becomes highly non-local in time. An increasingly long equilibration time is needed for the effect of the initial condition to disappear.
So for a system with long memory ($\rme^{-\lamg\tau}\approx 2$), what happens if, instead of an equilibrated feedback system, we start from $t_0=0$, i.e.~the particle sitting in the centre of the feedback potential? 
\begin{figure}[tb]
\centering
\includegraphics[width=0.99\columnwidth]{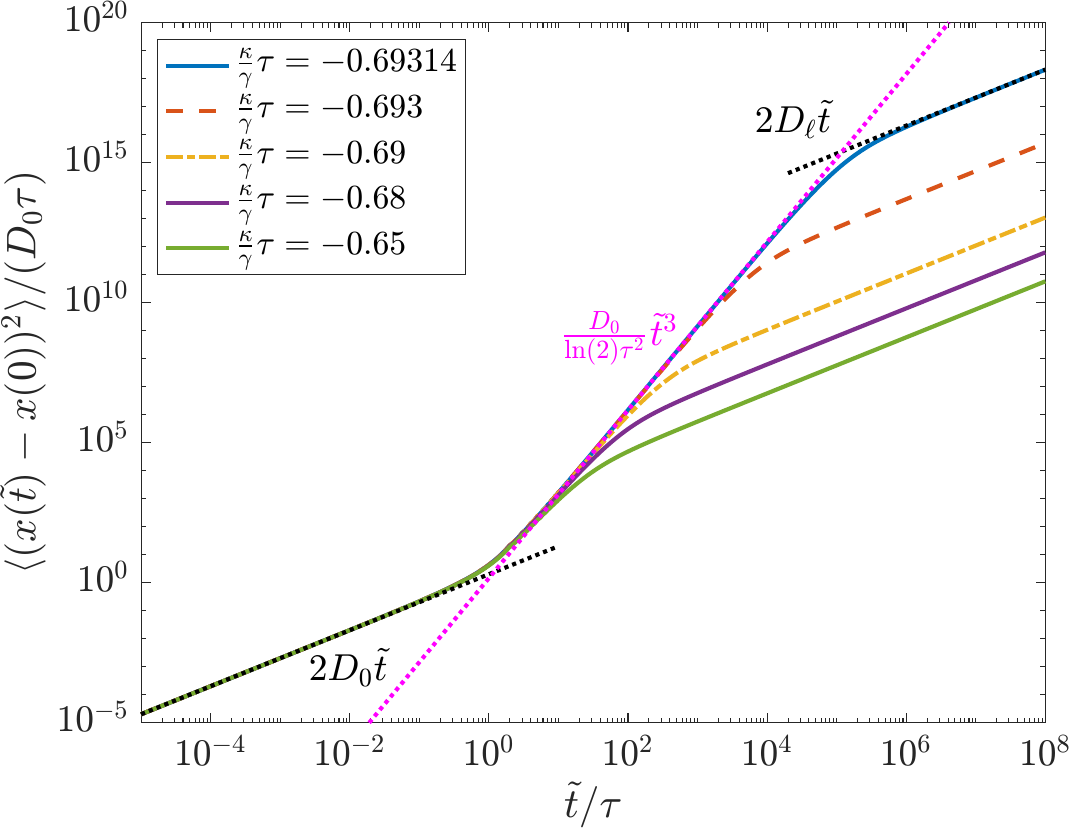}
\caption{Analytic result for the MSD in moderately strong repulsive harmonic feedback potentials of varying strength close to the divergence point $\rme^{-\lamg\tau}=2$ and starting with the particle in the potential centre (i.e.~$t_0=0$ fixed). The repulsive feedback potential, pushing the particle away from its past position, leads to an effective propulsion of the particle. Due to the growing (variance of the) particle velocity over time, the particle shows super-ballistic $\propto \tilde{t}^3$ at intermediate times. The long-time diffusion coefficient remains the same as for the equilibrated system. The magenta-colored line corresponds to the highest order term of the MSD given in eq.~(\ref{eq_MSD_log2}), i.e.~in the limit $\lamg\tau\to-\ln{2}$. [For reference: $\ln(2)\approx 0.69314718056$]}\label{fig_AnaMSD_rep_t0}
\end{figure}
In this case, the resulting MSD shows super-ballistic $\propto \ttil^3$ behaviour instead of ballistic motion, see Fig.~\ref{fig_AnaMSD_rep_t0}.
From eq.~(\ref{eq_MSD}), we find that for the MSD in the limit $\lamg\tau\to -\ln(2)$ the highest order in time is indeed given by
\begin{align}
   \lim_{\lamg\tau\to -\ln(2)}&\langle \left( x(\ts+\ttil)-x(\ts)\right)^2\rangle\propto
   \frac{D_0\tau}{\ln(2)}\tl^3 \propto \frac{D_0}{\ln(2)\tau^2}\ttil^3\label{eq_MSD_log2}
\end{align}
where we used $\tl=(\ttil+t_0-\DeltatE)/\tau$ and the limit values
\begin{equation}
    \lim_{\lamg\tau\to 2}c_n=-\left(\frac{1}{2}\right)^n\left(n+1\right)
\end{equation}
and
\begin{align}
\lim_{\lamg\tau\to-\ln(2)} A(i,j)&=-\frac{ 2^{-(j-i+1)} }{3}\,\left(i^3-3i^2j-9ij-7i-6j-6\right)\,.
\end{align}
Mathematically, the $\ttil^3$ appears from the $A(i,j)$ terms, specifically 
$A(\tl-1,\tl-1)$, $A(\tl-2,\tl-1)$, $A(\tl-1,\tl-2)$, $A(\tl-2,\tl-2)$ all scale with $\tl^3$ as their highest order.
Physically, the $\propto \ttil^3$ behaviour can be understood in the following way:
Starting from the coarse-grained particle velocity \mbox{$\bar{v}_k:=(x(k\tau)-x((k-1)\tau))/\tau$}, that follows from eq.~(\ref{eq_ansatz_xktau}) as
\begin{align}
\bar{v}_k&:=\frac{x(k\tau)-x((k-1)\tau)}{\tau}\nonumber\\
&=\frac{1}{\tau}\sum_{n=0}^{k-1}\left(\rme^{-\lamg\tau}-1\right)^n\rme^{-\lamg (k-n)\tau}\int_{((k-n-1)\tau}^{(k-n)\tau}\rmd t' \frac{\mathrm{f}(t')}{\gamma}\rme^{\lamg t'}
\end{align}
we find that the velocity average $\langle \bar{v}_k\rangle=0$ is zero, while its mean squared value changes as
\begin{align}
\langle\bar{v}_k^2\rangle&=\Big\langle\frac{ \left[ x(k\tau)-x((k-1)\tau)\right] ^2 }{\tau^2}\Big\rangle\nonumber\\
&=\frac{1}{\gamma^2\tau^2}\sum_{n=0}^{k-1}\sum_{m=0}^{k-1}\left(\rme^{-\lamg\tau}-1\right)^{n+m}\rme^{-\lamg (k-n)\tau}\rme^{-\lamg (k-m)\tau}\nonumber\\
&\quad\int_{(k-n-1)\tau}^{(k-n)\tau}\rmd t'\int_{(k-m-1)\tau}^{(k-m)\tau}\rmd t'' \underbrace{\langle \mathrm{f}(t')\mathrm{f}(t'')\rangle}_{2D_0\gamma^2\delta(t'-t'')}\rme^{\lamg (t'+t'')}\nonumber\\
&=\frac{2D_0}{\tau^2} \sum_{n=0}^{k-1}\left(\rme^{-\lamg\tau}-1\right)^{2n} \rme^{-2\lamg (k-n)\tau}\int_{(k-n-1)\tau}^{(k-n)\tau}\rmd t' \rme^{2\lamg t'}\nonumber\\
&=\frac{\gamma D_0}{\kappa \tau^2}\left(1-\rme^{-2\lamg \tau}\right) \sum_{n=0}^{k-1}\left(\rme^{-\lamg\tau}-1\right)^{2n} \,.
\end{align}
For long memory, i.e., close to the divergence point $\rme^{-\lamg\tau}=2$, the squared coarse-grained velocity is thus given by
\begin{align}
\langle\bar{v}_k^2\rangle
&\approx-\frac{3\gamma D_0}{\kappa \tau^2} \sum_{n=0}^{k-1}\left(1\right)=-\frac{3\gamma D_0}{\kappa \tau^2}\,k\label{eq_velsq}
\end{align}
and thus grows linearly in $k$ (where $k\approx t/\tau$), i.e.~$\propto t$, leading to a $\langle \bar{v}^2\rangle \ttil^2 \propto \ttil^3$ behaviour in the MSD. The super-ballistic behaviour is thus caused by the growing variance of the particle velocity after the feedback potential is introduced which is similar to the $\ttil^3$-scaling of an active particle in shear flow \cite{tenHagen2011}. Further, for appropriate choices of $\lamg$ and $t_0$, the particle shows both ballistic and super-ballistic motion with the proportionality in the MSD changing as $\ttil$ - $\ttil^2$ - $\ttil^3$ - $\ttil$, see Fig.~\ref{fig_AnaMSD_rep_t0_t3}. The $\ttil^2$ and $\ttil^3$ regimes correspond to motion at constant velocity, given by the particle velocity at $t_0$ according to eq.~(\ref{eq_velsq}), and the increase in velocity due to the long memory in the system.

\begin{figure}[tb]
	\centering
\includegraphics[width=0.99\columnwidth]{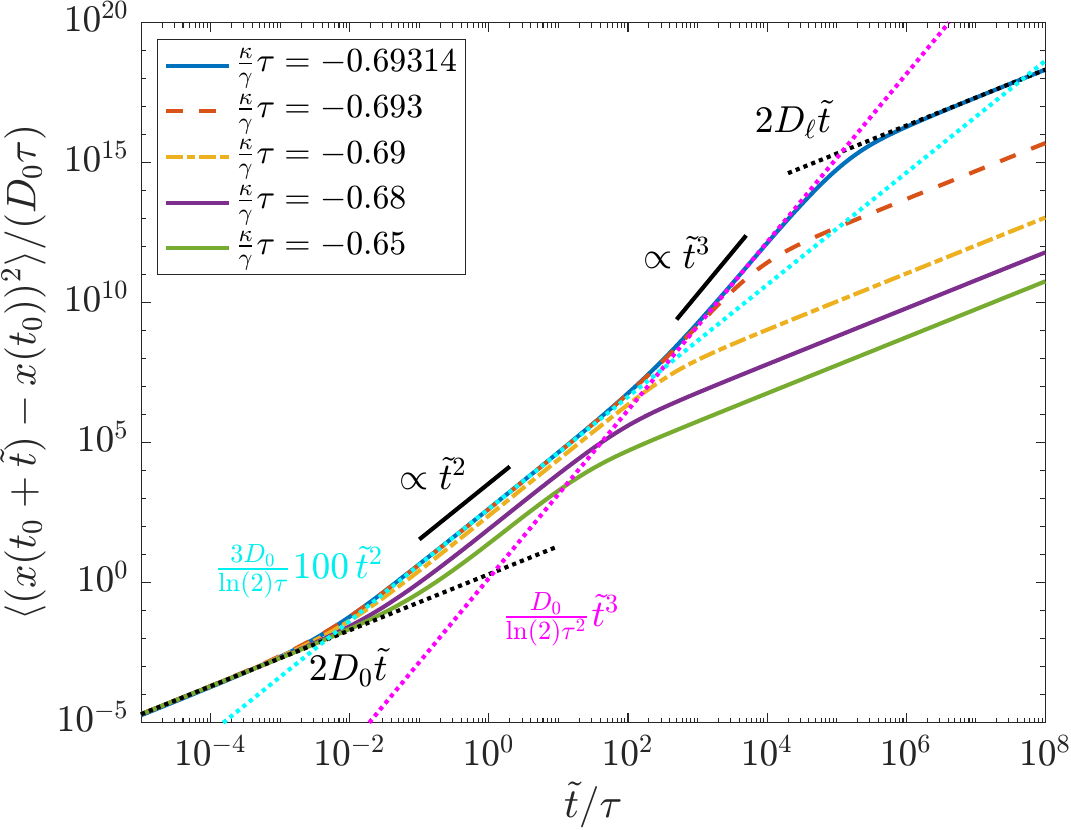}
\caption{Analytic result for the MSD in repulsive harmonic feedback potentials of varying strength starting from a not fully equilibrated feedback system (average taken over $t_0\in [100\tau, 101\tau)$ after first introduction of the feedback potential). The initial particle velocity (given by eq.~(\ref{eq_velsq}) after $100\tau$) leads to a $\propto \ttil^2$ regime. The cyan-colored line corresponds to a particle that travels at this constant velocity. The additional speed-up with time leads to a $\propto \ttil^3$ regime after multiple potential renewals. The magenta-colored line is the same as in Fig.~\ref{fig_AnaMSD_rep_t0}. [For reference: $\ln(2)\approx 0.69314718056$]}\label{fig_AnaMSD_rep_t0_t3}
\end{figure}

In our calculations, we have derived a full analytic solution for the trajectory a single particle in a harmonic feedback potential. We have shown that the feedback potential can lead to both reduced and increased long-time diffusion as well as particle propulsion at intermediate times. Furthermore, care must be taken when defining the mean squared displacement (MSD) as the system is not completely symmetric under shifts in time and strong memory effects can lead to the retaining of the initial condition for long times. In the following, we drop the tilde from $\ttil$ in the MSD, implying that the system is equilibrated long enough for the initial condition to be lost, and that the average over $t_0$ is executed properly such that only the relative time offset is relevant.

\subsection{Experimental realization and simulation results}
\label{sec:individual}

In the experiments, two types of feedback potentials are realized: First, an attractive harmonic potential is implemented. Second, instead of a repulsive harmonic potential (which is technically hard to realize in the experiments) we use an attractive annular ring potential. This potential pulls the particle towards a bright ring centered around its past position, thus leading to an \textit{effective repulsion} from its past position.\\

\subsubsection{Attractive harmonic potential}
An attractive harmonic potential with stiffness $\kappa>0$ is imposed onto the particle centered around its previous position. While the actual light beam has a Gaussian shape, the particle stays close to the potential centre such that the potential can be approximated as harmonic (Fig.\ref{fig:attractive}A). Due to the time delay $\tau$ of the feedback loop, the harmonic potential centre is located at the previous (rather than the actual) particle position and thus the particle tends to move back towards it (Fig.\ref{fig:attractive}B).

\indent The experimental setup was realized in three different ways: First, the spatial light modulator (SLM) was used as a mirror only and the Gaussian shape of the beam was exploited to create the desired light pattern. The beam location was controlled using galvanometer-mounted mirrors (GMM) (Fig.\ref{fig:attractive}C). Second, the SLM was used to create a harmonic potential of different stiffnesses $\kappa$ and the beam location was again controlled using the GMM (Fig.\ref{fig:attractive}D). Third, the SLM was used to create a harmonic potential with different $\kappa$ and to steer the beam (not shown in this figure, but discussed for other examples, see Figs.~\ref{fig:chains},\ref{fig:dimers}). All three realizations yielded equivalent results and perfectly agree with our analytic calculations.\\

\indent In particular, we measured the two-dimensional mean squared displacement (MSD) $\langle (\vect{r}(t+t_0)-\vect{r}(t_0))^2\rangle$, as a function of time $t$ (with the average taken over $t_0$) first for a static harmonic potential (no feedback) and then for the feedback case. The static harmonic potential case is used to calibrate the potential strength, i.e.~determine $\kappa$. By fitting the MSD to the known result $\langle (\vect{r}(t+t_0)-\vect{r}(t_0))^2\rangle=\frac{4D_0}{\kappa/\gamma}(1-\rme^{-\frac{\kappa}{\gamma}t})$ for a harmonic trap\cite{Lukic2007}, we obtain both the diffusion coefficient $D_0$ within the potential as well as the potential stiffness $\kappa$, see dotted and dashed lines in Figs.\ref{fig:attractive}C,D. 
Next, the feedback loop is introduced, i.e.~the same harmonic potential is applied with its centre at the position the particle was at a time $\tau = 0.2\mathrm{s}$ earlier. As predicted by our analytic calculations, at short times, $t < 0.1\mathrm{s}$, diffusive motion is observed; at intermediate times, $0.1\mathrm{s} < t < 3\mathrm{s}$, the delayed feedback results in oscillations; at long times, $t > 3\mathrm{s}$, diffusion is re-established with a significantly reduced diffusion coefficient. The long time diffusive behaviour then corresponds to the diffusive motion of the position of the potential centre. The experiments perfectly agree with our theoretical predictions (Fig.\ref{fig:attractive}C,D).\\

\subsubsection{Attractive annular (=effectively repulsive) potential}
\begin{figure*}[tbh]
	\centering
	\includegraphics[width=6in]{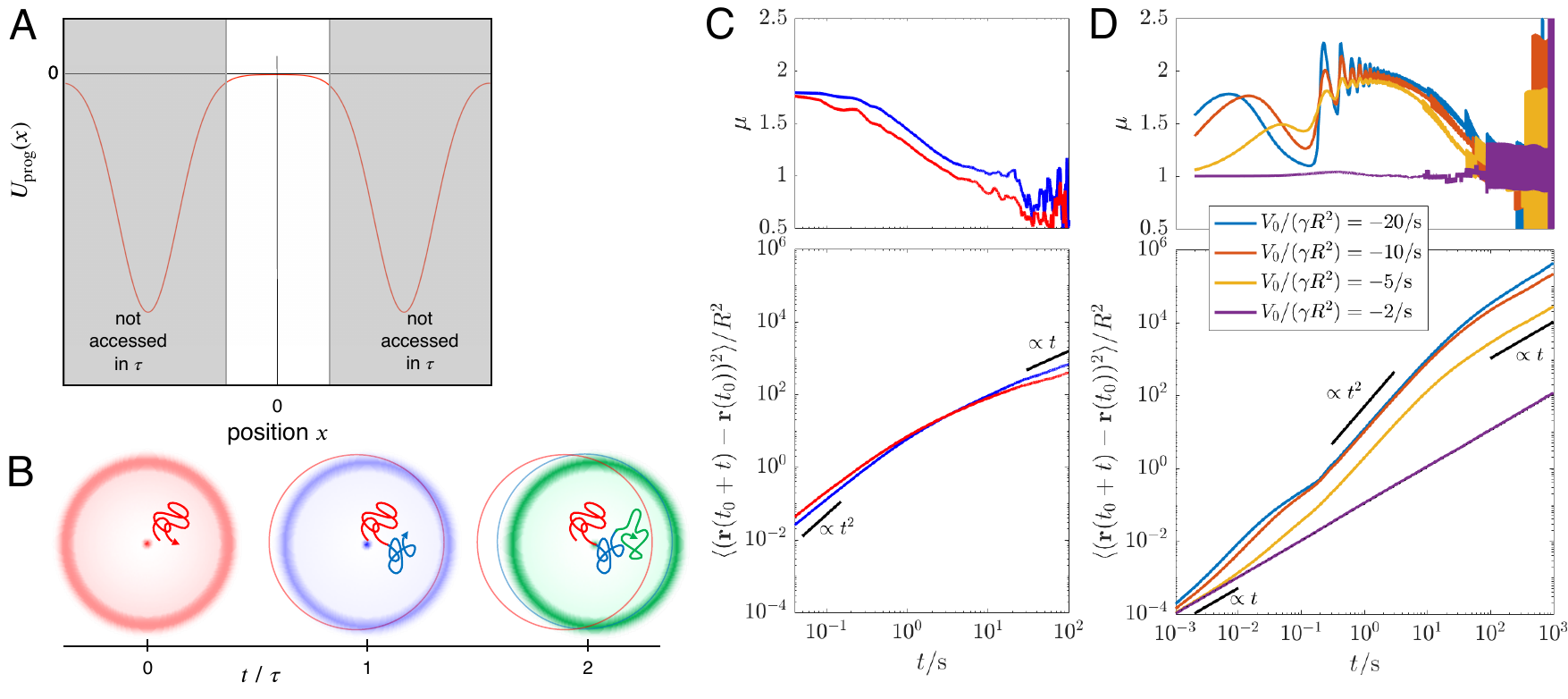}
	\caption{(A) One dimensional cut through the applied annular potential. (B) The potential is centered around the position that the particle occupied a time $\tau$ earlier. Trajectories within the potential are indicated by the same color.
		The potential is kept the same for a duration $\tau$ while the next potential is calculated. The starting point of the colored trajectory then becomes the centre of the next programmed potential. The repeated relocation of the potential continuously pushes the particle away from the centre. Additionally, due to the time delay $\tau$, the particle is typically off-centre in the refreshed potential. During two successive steps, the particle (as well as the potential) hence tends to advance in the same direction. (C,D) Mean squared displacement (MSD) and exponent $\mu$ as a function of time $t$ in the experiment (C) and in the simulations (D). The programmed feedback potential leads to a $\propto t^2$ behaviour at intermediate times ($\mu\approx 2$) while the particle dynamics is diffusive at both short and long times ($\mu\approx 1$). The experimental results show the transition from ballistic to diffusive motion. (The initial diffusive regime is not covered). The simulations reveal additional oscillatory behaviour with period of about $\tau$ resulting from the discretized potential update. [Experimental setup in (C): Potential created with the SLM, potential strength is slightly larger for the red curve; Simulation parameters for (D): $D_0=0.026R^2/\mathrm{s}$, $\tau=0.2\mathrm{s}$, $b^2=0.1R^2$, $r_\mathrm{ring}=1.4 R$, potential amplitude given in figure legend, MSD averaged over $20000\mathrm{s}$.]}
	\label{fig:repulsive}
\end{figure*}

Next, we applied an annular ring potential, see Fig.~\ref{fig:repulsive}A,B. Due to the form of the potential, the particle is pulled towards the ring surrounding its past position and is thus effectively pushed away from its past position. We use a ring potential of radius $1.4R$ (where $R$ is the particle radius). In this case, the circular minimum typically is not reached by the particle within the delay time $\tau$ before the potential is relocated (see Fig.~\ref{fig:repulsive}A,B). The light pattern is created using the SLM.

Three regimes of the MSD are observed (Fig.~\ref{fig:repulsive}C): At short times, the particle freely diffuses with diffusion coefficient $D_0$. 
These times are typically shorter than the delay time $\tau = 0.2$~s and hence are not recorded in the experiments (Fig.~\ref{fig:repulsive}C) but covered in the simulations (Fig.~\ref{fig:repulsive}D). 
At intermediate times, $0.03\mathrm{s} < t < 1\mathrm{s}$, the motion is superdiffusive, almost ballistic. This is also indicated by the dynamical exponent \mbox{$\mu = \rmd\ln{\langle  \left( \vect{r}(\ts+t)-\vect{r}(\ts)\right)^2 \rangle} \,/\,\rmd\ln{t}$}, which reaches a value of almost $2$ in the experiments. 
At these times, the repeated relocation of the potential continuously pushes the particle away from the potential centre causing an effective propulsion. As the particle is typically off-centre in the refreshed potential, it tends to proceed in the same direction during successive steps. This propulsion direction is retained for long times ($\approx10\mathrm{s}$). Since the particle also undergoes spatial diffusion, the propulsion direction slowly changes, such that at very long times diffusion is re-established, with a diffusion coefficient higher than the short-time one. The long-time diffusive motion can be understood as the diffusive motion of the potential centre. Again, the experimental findings qualitatively agree with simulations, in particular the transition from ballistic $\propto t^2$, to diffusive, $\propto t$ motion (Fig.~\ref{fig:repulsive}C,D). The simulations show additional oscillations of period $\tau$ in $\mu$ that are caused by the discretized potential updates: The effective propulsion is caused by the feedback force whose strength varies with the distance of the particle 
from its past position. The force thus varies periodically with period $\tau$.\\


\FloatBarrier
\section{Feedback-driven dynamics of groups of colloidal particles}\label{ch_feedback_multip}
In addition to individual particles, we also studied an ensemble of particles to explore their cooperative behaviour. Here, we program a potential landscape made up of annular rings centered around all previous particle positions.

\begin{figure}[tbh]
\centering
\includegraphics[width=0.8\columnwidth]{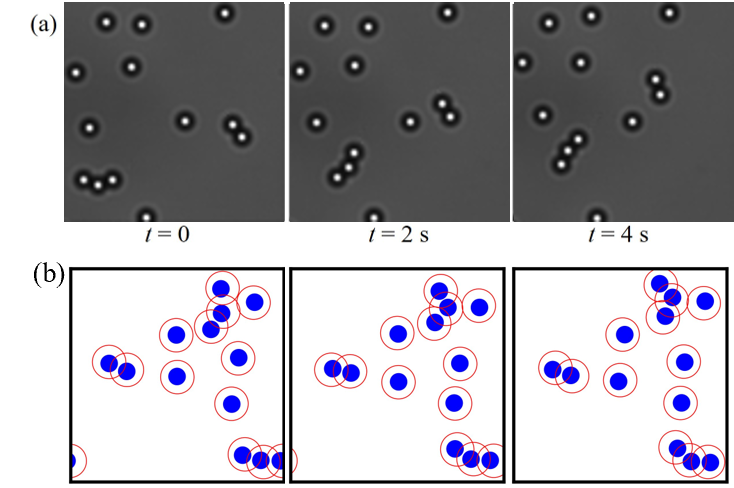}
\caption{Cooperative motion induced through programming. (a) Particles with an attractive annular potential with radius $2R$ and delay time $\tau = 0.2$~s assemble in `living' chain-like structures in the experiment. (b) Similar chain-like structures are obtained in the simulation. [Simulation parameters: $N=13$, $D_0=0.026R^2/\mathrm{s}$, $\tau=0.2\mathrm{s}$, $V_0/(\gamma R^2)=-0.35/\mathrm{s}$, $b^2=0.02R^2$, $r_\mathrm{ring}=2R$, $\rho=0.02/R^2$]}
\label{fig:chains}
\end{figure}

\begin{figure*}[bth]
	\centering
	\includegraphics[width=0.95\textwidth]{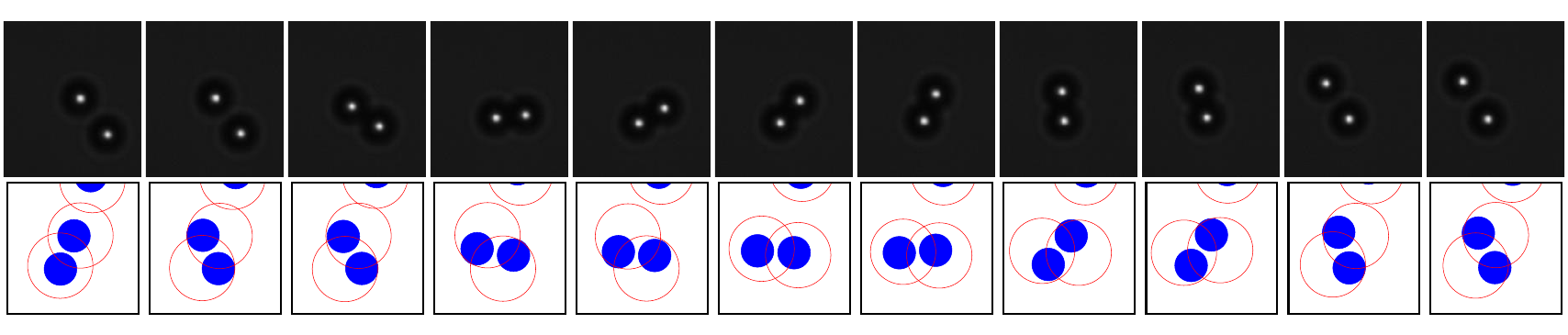}
	\caption{(Top) Translation and rotation of dimers assembled through the feedback potential. Attractive annular potential with radius $4R$ and delay time $\tau = 0.2$~s. The micrographs are taken in intervals of $0.2$~s. (Bottom) Simulation snapshots for an attractive annular potential. [Simulation parameters: $N=13$, $D_0=0.026R^2/\mathrm{s}$, $\tau=0.2\mathrm{s}$, $V_0/(\gamma R^2)=-20.0/\mathrm{s}$, $b^2=0.02R^2$, $r_\mathrm{ring}=2.0 R$, $\rho=0.02/R^2$]}
	\label{fig:dimers}
\end{figure*}

For multiple particles and no feedback potential in a dilute system, the particles are found in a disordered fluid state. By imposing an attractive interaction potential, however, ordered structures can be induced and the formation of ``living'' chains and small clusters (Fig.~\ref{fig:chains}) was observed. Through collisions, these chains grow and only very rarely break. Similar to the individual particles, the chains undergo superdiffusion.
In very dilute samples, dimers exist for long times before forming trimers. They can be observed to not only perform translational motion but also rotate and slightly change their separation (Fig.~\ref{fig:dimers}). Similar chains and ``rotators'' were observed in computer simulations of the system. It is remarkable that the directed motion results from an instability-like situation: using an initial fluctuation the system is set into directed motion, while the subsequent iterations amplify the stochastic feedback systematically.

\begin{figure*}[tbh]
\centering
\includegraphics[width=6.5in]{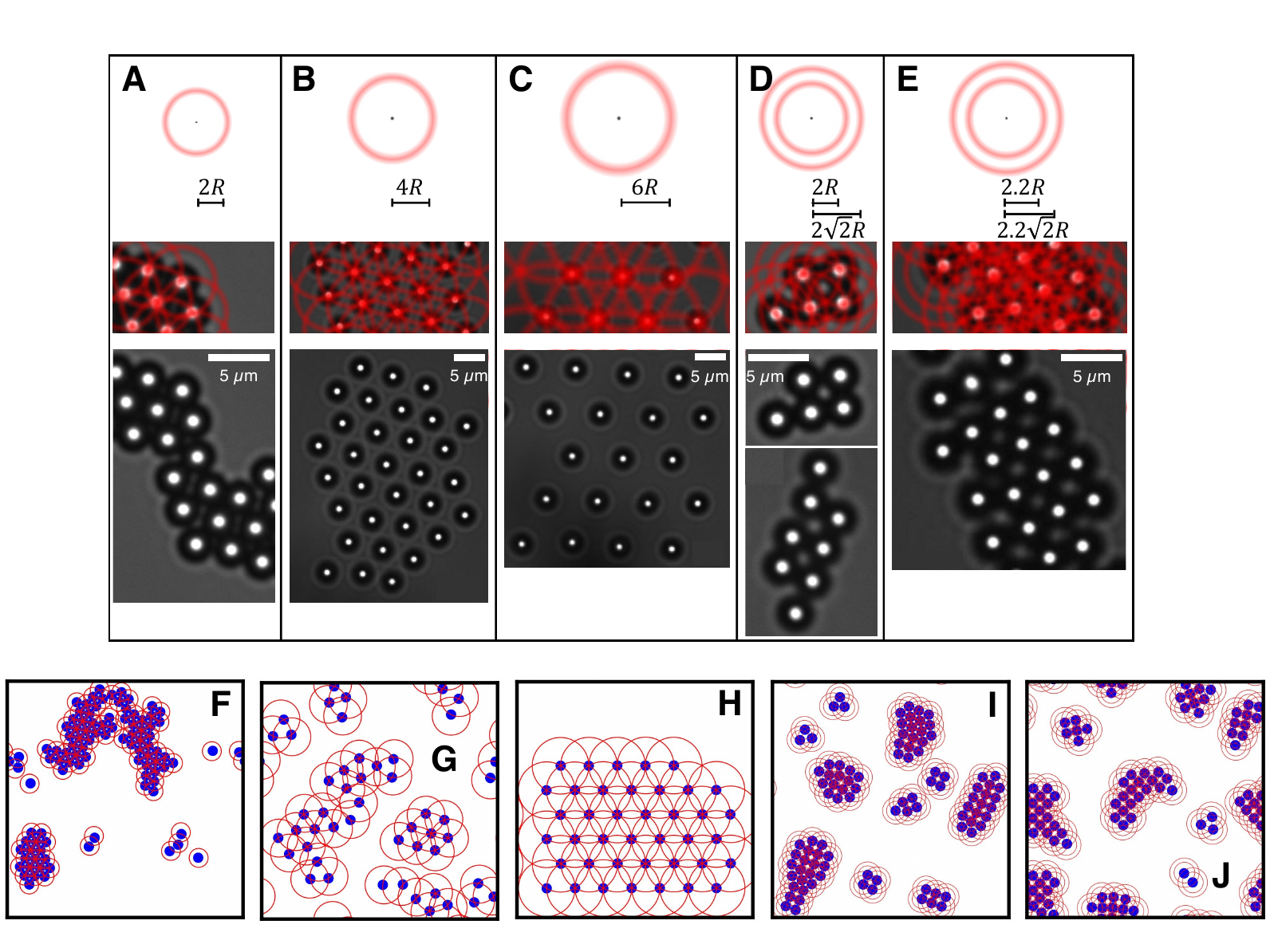}
\caption{Particle clusters formed for different parameters of the feedback potential. (A-E) Experimental results. (top) Schematic representation of the respective particle-particle interaction potential, which determines the total potential (middle) imposed on the particles (schematically represented in red), and the resulting particle arrangement (bottom). (F-H) Corresponding simulation snapshots. ``Single ring'' feedback potentials (A-C) lead to triangular lattices of corresponding spacing. ``Double ring'' feedback potentials (with ratio of radii $\sqrt{2}$) (D,E) can be used to stabilize a square lattice.
[Simulation parameters: $D_0=0.026R^2/\mathrm{s}$, $\tau=0.2\mathrm{s}$, 
(F-H) $V_0/(\gamma R^2)=-10.0/\mathrm{s}$, $b^2=0.02R^2$,
(I, J)$V_0/(\gamma R^2)=-5.0/\mathrm{s}$, $b^2=0.25R^2$, $N=100$, $\rho=0.04/R^2$,
(F) $N=100$, $\rho=0.04/R^2$, $r_\mathrm{ring}=2 R$, (G) $N=50$, $\rho=0.02/R^2$, $r_\mathrm{ring}=4 R$, (H) $N=40$, $\rho=0.016/R^2$, $r_\mathrm{ring}=6 R$,
(I)  $r_\mathrm{ring}=2 R$, $r_\mathrm{ring}=2\sqrt{2} R$, 
(J)  $r_\mathrm{ring}=2.2 R$, $r_\mathrm{ring}=2.2\sqrt{2} R$. All simulations were started from randomly placed particles except for (H) which was started from a triangular lattice.]}
\label{fig:arrangement}
\end{figure*}

Further, we programmed an attractive annular interaction potential with radius $2R$ and width $R/7$ and found crystallization of the particles into a triangular lattice (Fig.~\ref{fig:arrangement}A). The triangular arrangement optimises the particle separation (to the radius of the interaction potential) and maximizes the number of neighbours with this separation. During the assembly process, the particles are observed to rotate around each other (similar to the ``rotators'' in Fig.~\ref{fig:dimers}) thus keeping their ideal separation while more neighbours approach (similar to the chain formation in Fig.~\ref{fig:chains}). This separation, i.e.~the size of the crystal unit cell, can be tuned through the radius of the annular interaction potential (Fig.~\ref{fig:arrangement}A-C).\\

\indent
An interaction potential with two concentric attractive annuli of ratio $\sqrt{2}$ (e.g.~radii $2R$ (as before) and $2\sqrt{2}R$) and identical width $R/7$ (as before) leads to a square arrangement of the particles (Fig.~\ref{fig:arrangement}D). The larger distance along the diagonal of the unit cell is favoured by the second annulus which overcompensates the effect of the reduced number of neighbours and hence stabilises the square lattice.
Again, an increase in the radii of the annuli leads to larger unit cells (Fig.~\ref{fig:arrangement}E). The crystal formation was also found for computer simulations of the respective systems (Fig.~\ref{fig:arrangement}F-J).
Note that these crystal structures are not the result of a fixed template that favors certain spatial positions, but is a result of the self-organization of the particles according to the programmed interaction potential. The formed crystals are therefore invariant under translation and rotation and hence retain all possible zero energy modes.

\section{Conclusions}
We have introduced cybloids, i.e.~cybernetic colloids, whose interaction potential is computer-controlled through a programmed feedback potential. We have systematically and quantitatively investigated the characteristics of these cybloids, including their arrangement, dynamics and collective behaviour (Figs.~\ref{fig:attractive}, \ref{fig:repulsive}, \ref{fig:chains}, \ref{fig:dimers} and \ref{fig:arrangement}).

Here, we have shown that the dynamics of the cybloids can be tuned by the programmed potential and delayed feedback. Through the programming, a cybloid particle is made to diffuse on short and long times with the diffusion coefficients given by the particle's and the potential's diffusion coefficients, respectively while for intermediate times transient sub- or superdiffusion appears. This demonstrates that a broad range of dynamics is accessible and can be comprehensively investigated using this approach.
Further, for groups of particle, we have shown the formation of ``living'' chains and small clusters as well as crystal lattices. 
Crystal formation can be induced and the crystal structure and the unit cell size controlled. This is achieved through an externally imposed interaction potential between the particles, not a stationary particle-independent potential, i.e.~fixed template, for example, an array of tweezers. Programmed particles hence provide the opportunity to systematically and quantitatively investigate the relation between particle-particle interactions and particle arrangements, e.g. different crystal symmetries.
This example shows that programming can also be exploited to induce cooperative behaviour, similar to the recently reported implementation of specific cooperation rules \cite{Lavergne_2019}.

Beyond these examples, programming can be exploited to impose pair and multi-body interactions, to spur drifts, bias and swarming, to induce memory effects or to couple two otherwise unrelated samples and to ``clone'' samples. Many further dependencies, situations and couplings can be imagined. In any case, the properties of each particle can be prescribed and hence the sample is controlled on all relevant length scales, without the need to change or directly interfere with the sample nor the need that such a sample can actually be prepared. This ideally complements the already very popular possibility to observe and follow each particle \cite{Prasad2007,Royall2007,Lee2007,Jenkins2008,Allahyarov,Leocmach2013} and hence characterize the system on all relevant length scales. Therefore, cybloids open a huge range of new opportunities which can be exploited to create previously inaccessible model systems, explore the rational design of materials or mimic natural and industrially relevant situations.

\section*{Author Contributions}
DS performed the experiments, ST did the theoretical calculations and performed the simulations. SUE and HL supervised the project. DS, ST, SUE and HL discussed the results and contributed to writing the paper. 

\section*{Competing interest}
The authors declare no competing interests.

\section*{Acknowledgements}
We mourn the loss of our colleague Stefan Egelhaaf, who left us far too early. Many of the ideas in this paper originated from him. We are grateful for his creativity, his thoroughness, his persistence in pursuing questions and questioning data until they were consistent and understood, and for the many joint discussions that ultimately led to this publication. This work was financially supported by a joint project of SUE and HL of the German Research Foundation (DFG) with the project numbers EG 269/6 and LO 418/19. DS thanks Dennis Rohrschneider for his help with the experiments.

\FloatBarrier
\bibliography{cybloid_ref}

\end{document}